\documentclass[smallextended]{svjour3} 
\usepackage{graphicx}  
\usepackage{amsmath,amssymb}
\usepackage[titletoc]{appendix}
\usepackage{dsfont}
\usepackage{xcolor}
\usepackage{afterpage}
\usepackage{amsfonts}
\usepackage{color}
\usepackage[all]{xy}
\usepackage{mathtools}
\usepackage{tcolorbox}
\usepackage[nottoc]{tocbibind}
\usepackage{wasysym}

\usepackage{bbm}

\renewcommand{\1}{\mathbbm{1}}

\newcommand{\N}{\mathbb{N}}
\newcommand{\C}{\mathbb{C}}
\newtheorem{conj}{Conjecture}

\newcommand{\tr}{\operatorname{Tr}}
\newcommand{\id}{\mathds{1}}
\newcommand{\myinv}[1]{#1^{\scalebox{0.9}[1.0]{-}1}}

\newcommand{\beq}{\begin{equation}}
\newcommand{\eeq}{\end{equation}}
\newcommand{\beqa}{\begin{eqnarray}}
\newcommand{\eeqa}{\end{eqnarray}}
\newcommand{\ket} [1] {\vert #1 \rangle}
\newcommand{\bra} [1] {\langle #1 \vert}

\usepackage{tikz}
\usetikzlibrary{3d}
\usetikzlibrary{shapes,snakes}
\usetikzlibrary{decorations.pathreplacing,matrix,arrows.meta,decorations.markings,calc}

\tikzset{
  every picture/.style = {
    baseline={([yshift=-.5ex]current bounding box.center)}, 
    scale=1.2,
    transform shape,
    font=\scriptsize
  }
}

\tikzset{
  tensor/.pic={
    \begin{scope}[canvas is zx plane at y=0]
      \filldraw (0,0) circle (0.07);
    \end{scope}
    \draw (0,0,0)--(0,0.2,0);
    
  }
}

\tikzset{
  tensord/.pic={
    \begin{scope}[canvas is zx plane at y=0]
      \filldraw (0,0) circle (0.07);
    \end{scope}
    \draw (0,0,0)--(0,-0.2,0);
    
  }
}
\tikzset{
  tensorud/.pic={
    \begin{scope}[canvas is zx plane at y=0]
      \filldraw (0,0) circle (0.07);
    \end{scope}
    \draw (0,0.2,0)--(0,-0.2,0);
    
  }
}

\tikzset{
  3dpepsres/.pic={
    \begin{scope}[canvas is zx plane at y=0]
      \draw (-0.5,0)--(0.5,0);
      \draw (0,-0.4)--(0,0.4);
      \filldraw (-0.07,-0.07) rectangle (0.07,0.07); 
    \end{scope}
    \draw (0,0,0)--(0,0.16,0);
  }
}
\tikzset{
  3dpepsdown/.pic={
    \draw (0,0,0)--(0,-0.2,0);
    \begin{scope}[canvas is zx plane at y=0]
      \draw (-0.5,0)--(0.5,0);
      \draw (0,-0.4)--(0,0.4);
      \filldraw (0,0) circle (0.07);
    \end{scope} 
  }
}
\tikzset{
  hopf/.style={
    line cap=round,
    line width=1.5mm,
  },
  epsilon/.style={
   draw = red,
   thin,
   densely dotted, 
   rounded corners,
   fill opacity=0.2,
   fill=red,
  },
  every picture/.style = {
    baseline={([yshift=-.5ex]current bounding box.center)}, 
    font=\scriptsize
  }  
}

\tikzset{
  pepsGisopen/.pic={
\begin{scope}[canvas is zx plane at y=0]
 \draw (0,0.07) --(0,0.3);
 \draw (0,-0.07) --(0,-0.3);
 \draw (-0.1,0) --(-0.4,0);
  \draw (0.1,0) --(0.4,0);
 \end{scope}
}}

\tikzset{
  tensor/.style={
    inner sep = 0.055cm,
    shape = circle,
    draw,
    canvas is zx plane at y=0,
    fill
  },
    tensorr/.style={
    inner sep = 0.045cm,
    shape = circle,
    draw,
    fill
  },
    tensorB/.style={
    inner sep = 0.065cm,
    shape = rectangle,
    draw,
    canvas is zx plane at y=0,
    fill
  },
  every picture/.style = {
    baseline={([yshift=-.5ex]current bounding box.center)}, 
    transform shape,
    font=\scriptsize
  }
}

\tikzset{
  every picture/.style = {
    baseline={([yshift=-.5ex]current bounding box.center)}, 
    scale=1.2,
    transform shape,
    font=\scriptsize
  }
}

\tikzset{
  3dpeps/.pic={
    \begin{scope}[canvas is zx plane at y=0]
      \draw (-0.5,0)--(0.5,0);
      \draw (0,-0.4)--(0,0.4);
      \filldraw (0,0) circle (0.07);
    \end{scope}
    \draw (0,0,0)--(0,0.17,0);
  }
}

\tikzset{
  O2/.pic={
\filldraw[rounded corners=.05cm] (-0.4,-0.05) rectangle (0.4, 0.05);
  \draw (-0.3,0.2)--(-0.3,-0.2);
    \draw (0.3,0.2)--(0.3,-0.2);
     }
}
\tikzset{
  O2p/.pic={
\filldraw[rounded corners=.02cm] (-0.2,-0.025) rectangle (0.2, 0.025);
  \draw (-0.15,0.1)--(-0.15,-0.1);
    \draw (0.15,0.1)--(0.15,-0.1);
     }
}
\tikzset{
  O2SVD/.pic={
\draw (-0.35,0)--(0.15,0);     
  \draw (-0.3,0.12)--(-0.3,-0.12);
    \draw (0.1,0.12)--(0.1,-0.12);
      \filldraw[draw=black, fill=blue] (-0.35,-0.05) rectangle (-0.25,0.05);
        \filldraw[draw=black, fill=purple] (0.05,-0.05) rectangle (0.15,0.05);
     }
}
    
 \tikzset{
  3dpepsp/.pic={
    \begin{scope}[canvas is zx plane at y=0]
      \draw (-0.3,0)--(0.3,0);
      \draw (0,-0.25)--(0,0.25);
      \filldraw[draw=black,fill=blue] (0,0) circle (0.07);
    \end{scope}
  }
}

\tikzset{
  pepsplus/.pic={
      \draw (0,0,-2)--(0,0,2);
      \draw (-2,0,0)--(2,0,0);
      \foreach \x/\z in {0/-1.5,0/0,-1.5/0,1.5/0,0/1.5}{
        \pic at (\x,0,\z) {3dpeps};
      }
  }
}
\tikzset{
  pepsplusres/.pic={
      \draw (0,0,-2)--(0,0,2);
      \draw (-2,0,0)--(2,0,0);
      \foreach \x/\z in {0/-1.5,0/0,-1.5/0,1.5/0,0/1.5}{
        \pic at (\x,0,\z) {3dpepsres};
      }
  }
}

\tikzset{
  pepsplusrescontrac/.pic={
      \draw (0,0,-2)--(0,0,2);
      \draw (-2,0,0)--(2,0,0);
      \foreach \x/\z in {0/-1.5,0/0,-1.5/0,1.5/0,0/1.5}{
        \pic at (\x,0,\z) {3dpepsres};
      }
       \foreach \c/\x in {zy plane at x/-1.5, zy plane at x/1.5,xy plane at z/-1.5,xy plane at z/1.5}{
        \begin{scope}[canvas is \c=\x]
          \draw (-0.5,0) rectangle (0.5,-0.15);
        \end{scope}
      }
  }
}

\tikzset{
  pepsplusdown/.pic={
      \draw (0,0,-2)--(0,0,2);
      \draw (-2,0,0)--(2,0,0);
      \foreach \x/\z in {0/-1.5,0/0,-1.5/0,1.5/0,0/1.5}{
        \pic at (\x,0,\z) {3dpepsdown};
      }
  }
}

\tikzset{
PAplus/.pic={
 \foreach \x/\z in {-2/0,-1.5/-0.5,-1.5/0.5, 1.5/0.5,1.5/-0.5,2/0}{
          \draw (\x,-0.15,\z)--(\x,0,\z);
        }
        \pic at (0,-0.15,0) {pepsplusdown};
        \pic at (0,0,0) {pepsplus};
        \foreach \x/\z in {-0.4/1.5, -0.4/-1.5,0/-2, 0/2, 0.4/-1.5, 0.4/1.5}{
          \draw (\x,-0.15,\z)--(\x,0,\z);
        }
 }
}

\tikzset{
  3dpepsdownp/.pic={
    \begin{scope}[canvas is zx plane at y=0]
      \draw (-0.3,0)--(0.3,0);
      \draw (0,-0.25)--(0,0.25);
      \filldraw[draw=black,fill=blue] (0,0) circle (0.07);
    \end{scope} 
  }
}

\tikzset{
  3dGisopeps/.pic={
\draw (-0.1,0,0)--(-0.1,0.4,0);
       \begin{scope}[canvas is zx plane at y=0]
          \draw (0.1,0)--(0.5,0);
        \draw (-0.5,0)--(-0.1,0);
       \filldraw (0.3,0) circle (0.04);
        \filldraw (0,-0.25) circle (0.04); 
    \end{scope} 
    
     \begin{scope}[canvas is zx plane at y=0.4]
     \filldraw (-0.3,0) circle (0.04);
     \filldraw (0,0.25) circle (0.04);
       \draw (-0.5,0)--(-0.1,0);
      \draw[preaction={draw, line width=1pt, white}] (0.1,0)--(0.5,0);
    \end{scope} 
     \draw (0,0,0.1)--(0,0.4,0.1);
\draw (0,0,-0.1)--(0,0.4,-0.1);
       \draw[preaction={draw, line width=1pt, white}] (0.1,0,0)--(0.1,0.4,0);
         \begin{scope}[canvas is zx plane at y=0]
          \draw (0,0.1)--(0,0.4);
       \draw (0,-0.4)--(0,-0.1);
          \end{scope} 
                  \begin{scope}[canvas is zx plane at y=0.4]
          \draw (0,0.1)--(0,0.4);
       \draw (0,-0.4)--(0,-0.1);
          \end{scope} 

  }
}

\begin{document}

\title{Mathematical open problems in Projected Entangled Pair States}

\author{J. Ignacio Cirac \and Jos\'e Garre-Rubio \and David P\'{e}rez-Garc\'{i}a}

\institute{J. Ignacio Cirac
\at Max-Planck-Institut fur Quantenoptik, Hans-Kopfermann-Strasse 1, D-85748 Garching, Germany
\and
Jos\'e Garre-Rubio \and David P\'{e}rez-Garc\'{i}a
              \at Departamento de An\'alisis Matem\'atico y Matem\'atica Aplicada, Universidad Complutense de Madrid, 28040 Madrid, Spain
           \and ICMAT, C/ Nicol\'as Cabrera, Campus de Cantoblanco, 28049 Madrid, Spain
}
\maketitle
  
\begin{abstract}
Projected Entangled Pair States (PEPS) are used in practice as an efficient parametrization of the set of ground states of quantum many body systems. The aim of this paper is to present, for a broad mathematical audience, some mathematical questions about PEPS.

\end{abstract}

\section{Introduction} 
Tensor network states play a prominent role in the rigorous study of central results in the theory of quantum many body systems -see \cite{reviewPEPS} for a complete review. In particular, PEPS capture the relevant physics in the low energy sector of local interacting systems. Then, the study of these systems is translated into the formal PEPS framework where different mathematical techniques have been developed. Fruitful advances in this field have arrived thanks to new connections that have been established with different areas in mathematics. 
The aim of this paper is to present some mathematical problems related to PEPS, providing the necessary background and motivation for them. The target audience is the general mathematical community, so some very basic notions on quantum physics are first introduced for the sake of completeness.
 
\subsection{Basic notions on quantum physics}

Quantum mechanics was developed in the middle 20's and, since then, it has become the underlying description of any physical model with the exception, yet, of gravitation. 
The generality of quantum mechanics as such universal framework came from its formulation as four axioms or postulates. These postulates provide the mathematical framework that has to be followed by any system and then one has to particularize them for the particular physical setup under consideration. The postulates are the following:\\

\paragraph{Systems and states.} This postulate provides the frame where the physical objects are placed; the system is described by a Hilbert space $\mathcal{H}$, which we suppose in this paper always finite-dimensional, so $\mathcal{H}=\mathbb{C}^d$, and the state is represented as a unit vector in that space.
We will use the standard notation in quantum mechanics, introduced by Dirac \cite{Dirac} and named  {\it bra-ket notation}, where (column) vectors $v \in \mathcal{H}=\mathbb{C}^d$ are denoted as \emph{kets} $\ket{v}$.  The adjoint (transpose conjugate) of $\ket{v}$ is denoted by the \emph{bra} $\bra{v}$, which can be understood just as a row vector in $\mathbb{C}^d$. Hence, the scalar product between $\ket{u}$ and $\ket{v}$ is just the product $\langle u |v \rangle$, and a rank-one operator adopts the form of the product $\ket{u}\bra{v}$.

  The simplest example, but extremely relevant, is $\mathcal{H}=\mathbb{C}^{2}$. It is known as a qubit system, which is the quantum analog of a bit, and the canonical basis is usually denoted as $\{ \ket{0},\ket{1}\}$. Instances of qubit systems are the spin of an electron or the polarization of a photon. Quantum mechanics also allows for the description of not completely known states; it is a probabilistic theory. These states are represented as \emph{density} matrices $\rho= \sum_i p_i \ket{\psi_i} \bra{\psi_i}$, called mixed states, where $p_i \ge 0$ represents the probability for the system to be in the state $\ket{\psi_i}$ so they fulfill $\sum_i p_i=1$. The (pure) state $\ket{\psi}$ is just represented as $\ket{\psi}\bra{\psi}$ in the density matrix formalism. 

\paragraph{Measurements.} This postulate describes the way in which quantum measurements are implemented and how they affect the measured system. The magnitude to measure is represented by a hermitian operator $\mathcal{O}$ called observable, in the case of projective measurements (see Ref. \cite{NielsenChuang} for the general case). The average value or the expectation value of $\mathcal{O}$ in the system described by the mixed state $\rho$ is $\langle \mathcal{O} \rangle_\rho=\tr[\mathcal{O}\rho]$ (which coincides with  $\bra{\psi} \mathcal{O} \ket{\psi}$ for a pure state $\ket{\psi}$).

\paragraph{Multiple systems.} The space associated to a composite system is mathematically represented by the tensor product of the components; $\mathcal{H}=\mathcal{H}_1\otimes\mathcal{H}_2 \otimes \cdots \otimes \mathcal{H}_N$. If we have $N$  systems each of them in the state $\ket{\psi_i}\in \mathcal{H}_i$ the global state is $\ket{\psi_1}\otimes\ket{\psi_2}\otimes \cdots \otimes \ket{\psi_N}\in \mathcal{H}$. However, there are also states that cannot be written in a tensor product form, these are called entangled states. Let us consider an example, a two qubit system $\mathcal{H}=\mathbb{C}^{2}\otimes\mathbb{C}^{2}$  in the state $\ket{\phi}=(\ket{0}\otimes\ket{1}+\ket{1}\otimes \ket{0})/\sqrt{2}$ cannot be written as $\ket{\phi}= \ket{a}\otimes\ket{b}$. This property is known as entanglement and it is believed to be the one endowing quantum mechanics its complexity. We will simplify the notation for tensor products writing $\ket{a}\ket{b}$ or $\ket{ab}$ instead of $\ket{a}\otimes\ket{b}$, so $\ket{\phi}\equiv (\ket{01}+\ket{10})/\sqrt{2}$.  \\

\paragraph{Evolution.} A quantum system changes with time according to a unitary transformation: $\ket{\psi(t_1)}=U(t_1,t_0)\ket{\psi(t_0)}$. The infinitesimal form of such evolution is described by the Schr\"{o}dinger equation:
\begin{equation}
i\hbar \frac{d\ket{\psi}}{dt }=H\ket{\psi},
\end{equation}
where $\hbar$ is Planck's constant and $H$ is the self-adjoint operator known as the Hamiltonian of the system.
 The Hamiltonian is the observable that measures the energy of the system. Since we are in the finite dimensional case, we can write the spectral decomposition of the Hamiltonian as $H=\sum_i E_i \ket{e_i}\bra{e_i}$, where the eigenvectors $\ket{e_i}$ are called energy eigenstates and $E_i$ is the energy of such state. The eigenstate corresponding to the smallest energy, $E_0\equiv {\rm min}\{E_i\}$, is known as the ground state (GS) and the other eigenstates are called excited states. The difference between the two smallest eigenvalues (energy levels) is known as spectral gap, or just gap, of the Hamiltonian and plays a fundamental role in many problems.\\

The Hamiltonian plays then a fundamental role in the description of a system. But its study encounters two main difficulties. On the one hand, this operator has to be deduced from the physics of the problem --the interaction between the parties among other considerations-- which is not a simple task. On top of that, the Hamiltonian obtained in this way would be in general very complex. To simplify the task, effective Hamiltonians are defined that aim to capture the relevant features of the system. Then, effective models are proposed to describe the low energy sector of the problem,  where the relevant quantum behaviors are expected to appear. On the other hand, when one has the effective model, it has to be solved. That is, the ground state and the low-energy excitations of the Hamiltonian have to be found  together their energies. Since we are interested in many body physics, that is, when a large number of parties is considered, we have to solve the problem in a \emph{huge} Hilbert space. Specifically the dimension of the total Hilbert space grows exponentially with the number of parties on it because of the inherent tensor product structure. This means that to describe any (entangled) state an exponential number of parameters is needed, which makes the task intractable. For example $N$ qubit systems are described by the Hilbert space ${\mathbb{C}^{2}}^{ \otimes N}$, so the dimension of the full space is $2^N$. But the naive fact that the full Hilbert space of \emph{any} quantum system grows exponentially with the number of parties on it is not an unavoidable obstruction since we are going to impose some restrictions on the systems of interest in the next subsection.

\subsection{Setup}

The systems that we consider are placed on one-dimensional or two dimensional finite size lattices $\Lambda$ where each vertex $v\in\Lambda$ represents a subsystem. The Hilbert space of each subsystem $\mathcal{H}_v$ is finite dimensional and isomorphic to $\mathbb{C}^{d_v}$. Then the total Hilbert space is:
$$\mathcal{H}_\Lambda=\bigotimes_{v\in \Lambda} \mathcal{H}_v. $$
We will focus on square lattices, so the one-dimensional case are just segments or rings of length $L$ for open boundary conditions or periodic boundary conditions respectively. In 2D we will consider an $L\times L$ square lattice. Therefore, for periodic boundary conditions  the system is placed on a torus ($L$ will correspond to the lattice size). 

The main assumption we will impose is that the interactions of the Hamiltonian are \emph{local}. This is motivated by the physical nature of the interactions: 
\begin{definition}
$H$ is a locally interacting Hamiltonian if it can be written as follows
$$H=\sum_i h_i\otimes \id_{\rm rest},$$ 
where every $h_i$ acts only in $\bigotimes_{v\in \Omega_i } \mathcal{H}_v$ and $\Omega_i$ is a connected sublattice of $\Lambda$ with $|\Omega_i|\le C$ ($C$ a constant independent of $i$ and $|\Lambda|$).
\end{definition}

We will further assume that the system is translationally invariant, meaning that $\mathcal{H}_v=\mathcal{H}_{v'}$ for all $v,v' \in \Lambda$, $\mathcal{H}_\Lambda= \mathcal{H}^{\otimes |\Lambda|}_v$, where $|\Lambda|$ is the total number of vertices, and the local terms $h\equiv h_i$ of $H$ are the same operators acting on translated sublattices. This implies that a given interaction $h$ defines the Hamiltonian $H$ for any lattice size.  

This allows us to define the limit when the system size grows to infinity (usually called thermodynamic limit). In particular, we can define the key notion of {\it gapped Hamiltonians}.

\begin{definition}
A family of Hamiltonians $H^{[L]}$ is gapped, where $L$ denotes the system size, if 
$$\Delta:=\liminf_{L\rightarrow \infty} \left(E_1^{[L]}-E_0^{[L]}\right)>0.$$
If this is the case $\Delta$ is called the gap of the system. 
\end{definition}
Note that for a finite Hilbert space the spectrum is discrete and then gapped, so the relevant information is how the gap behaves when the system size (and hence the Hilbert space dimension) grows to infinity.\\

 The key observation here is that  ground states of locally interacting gapped Hamiltonians have a very restrictive pattern of entanglement. Then, the states satisfying this pattern correspond to the subset of the full Hilbert space we are interested in. To describe and characterize this pattern let us introduce a measure of entanglement called entanglement entropy. Given a state $\ket{\psi}\in \mathcal{H}_A\otimes \mathcal{H}_{A^c}$ the reduced density matrix of the subsystem $A\subset \Lambda$ is defined as the partial trace on the complementary of $A$ in $\ket{\psi}$:
$$\rho_A=\tr_{A^c}\left[ \ket{\psi}\bra{\psi}\right],$$
where the partial trace is defined as the unique linear map fulfilling $$\tr_{A^c}\left[ \ket{a}\bra{b} \otimes \ket{u}\bra{v} \right]= \ket{a}\bra{b} \langle v\ket{u}$$ for $\ket{a},\ket{b}\in \mathcal{H}_A$, $\ket{u},\ket{v}\in \mathcal{H}_{A^c}$. The entanglement entropy of the subsystem $A$ is defined as follows
$$S_A( \ket{\psi})\equiv S_{VN}(\rho_A)= -\tr[\rho_A\log(\rho_A)],$$
where $S_{VN}$ is the von Neumann entropy. We will now recall some basic properties of the entanglement entropy that can be found in e.g. \cite{NielsenChuang}. The entanglement entropy of $A$ is equal to the one of $A^c$. For product states, i.e states that can be written as $ \ket{\psi}= \ket{\phi_A}\otimes  \ket{\sigma_{A^c}}$,  the entanglement entropy is zero. The entanglement entropy is bounded by the logarithm of the dimension of the Hilbert space where $A$ lives, $S_A( \ket{\psi})\le \log|\mathcal{H}_A|\propto |A|$. In fact this maximum rate, a scaling with the volume, is the typical behavior of a random state \cite{Hayden}. But for ground states of locally interacting gapped Hamiltonians the entanglement entropy of a subsystem is expected to scale as the boundary of the region:
$$S_A( \ket{\psi})\propto \log|\mathcal{H}_{\partial A}|\propto |\partial A|.$$
This is known as the Area Law Conjecture. It has been proven for one dimensional systems \cite{Hastings07}, \cite{Arad2013} and for some higher dimensional cases \cite{Hamza}, \cite{Masanes}. See \cite{Eisert} for a review.

The area law seems to be the characteristic property of ground states of locally interacting gapped Hamiltonian so the following question arises naturally: does there exist a tractable parametrization of the set of states fulfilling an area law? The answer is given by the so-called tensor network states, which by construction fulfill such entanglement pattern.

\subsection{Tensor Network States}\label{sec:TNS}
Tensor network states are multi-partite states on a lattice constructed with the contraction of local tensors placed on the vertices. A tensor is a vector $A\in \mathbb{C}^{d_1}\otimes \cdots \otimes\mathbb{C}^{d_r}$ where each element of the tensor product is called index, and $r$, the number of indices, is the rank of the tensor. The tensor product of two tensors $A\in \mathbb{C}^{d_1}\otimes \cdots \otimes\mathbb{C}^{d_r}$ and $B\in \mathbb{C}^{d'_1}\otimes \cdots \otimes\mathbb{C}^{d'_{r'}}$ is the tensor $A\otimes B\in\mathbb{C}^{d_1}\otimes \cdots \otimes\mathbb{C}^{d_r}\otimes \mathbb{C}^{d'_1}\otimes \cdots \otimes\mathbb{C}^{d'_{r'}}$ with rank $r+r'$. We define the contraction of two indices $i$ and $j$ with $d_i=d_j$ as the map:
\begin{align*}
\delta: \; & \mathbb{C}^{d_i}\otimes \mathbb{C}^{d_j} \to \mathbb{C}  \notag \\
 & \;\; \ket{\alpha \beta} \;\; \longmapsto \;\delta_{\alpha,\beta}  \notag,
\end{align*}
and extended by linearity. Then, the contraction of two indices of $A$ is carried out by acting with $\delta$ on those indices and with the identity on the rest of them. The resulting tensor $\delta\otimes \id_{\rm rest} (A)$ has rank $r-2$. We will use the standard graphical notation of tensors, where they are shapes with legs attached,  each of them representing an index: 
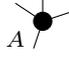
\begin{equation*}
\begin{tikzpicture}
\filldraw (0,0) circle (0.1);
  \draw (0,0)--(-0.1,-0.3);
  \draw (0,0)--(0,0.3);
     \draw (0,0)--(-0.3,0.2);
     \draw (0,0)--(0.3,0.1);
     \node at (-0.3,-0.2) {$A$};
 \end{tikzpicture}
 \end{equation*}
The contraction of two indices is represented as a line connecting the legs of the corresponding indices, thus $\delta\otimes \id_{\rm rest} (A)$ and the contraction between indices of different tensors $A$ and $B$ is  represented graphically as:
\begin{equation*}
\delta\otimes \id_{\rm rest} (A) \equiv
\begin{tikzpicture}
\filldraw (0,0) circle (0.1);
  \draw (0,0)--(-0.1,-0.3);
  \draw (0,0)--(0,0.3);
     \draw (0,0)--(-0.3,0.2);
     \draw (0,0)--(0.3,0);
     \node at (-0.3,-0.2) {$A$};
      \draw  (0.3,0) to [out=0, in=90] (0,0.3);
 \end{tikzpicture},
 \;\; \delta\otimes \id_{\rm rest} (A\otimes B) \equiv
 \begin{tikzpicture}
\filldraw (0,0) circle (0.1);
  \draw (0,0)--(-0.1,-0.3);
  \draw (0,0)--(0,0.3);
     \draw (0,0)--(-0.3,0.2);
     \draw (0,0)--(0.3,0);
     \node at (-0.3,-0.2) {$A$};
     \filldraw (1,0) circle (0.1);
  \draw (1,0)--(0.9,-0.3);
  \draw (1,0)--(1.1,0.3);
     \draw (1,0)--(0.7,0.2);
     \draw (1,0)--(0.7,-0.3);
     \node at (1.3,-0.3) {$B$};
     \draw  (0.3,0) to [out=0, in=-135] (0.7,-0.3);
 \end{tikzpicture}.
 \end{equation*}
The simplest examples of tensors are vectors and matrices:
\begin{equation*}
\ket{i}\equiv 
\begin{tikzpicture} [baseline=-1mm]
\filldraw (0,0) circle (0.08);
 \node at (-0.2,0.2) {$i$};
  \draw (0,0)--(-0.4,0);
 \end{tikzpicture}
\; ,\;\; A\equiv
 \begin{tikzpicture}[baseline=-1mm]
 \node at (0,-0.25) {$A$};
\filldraw (0,0) circle (0.1);
  \draw (-0.4,0)--(0.4,0);
 \end{tikzpicture}
 \in \mathbb{C}^D\otimes \mathbb{C}^D\cong \mathcal{M}_D,
 \end{equation*}
 so the multiplication of a vector by a matrix is
 \begin{equation*}
A\ket{i}\equiv 
\begin{tikzpicture}
\filldraw (0.4,0) circle (0.08);
\filldraw (0,0) circle (0.1);
 \node at (0.2,0.2) {$i$};
  \node at (0,-0.25) {$A$};
  \draw (-0.4,0)--(0.4,0);
 \end{tikzpicture},
 \end{equation*}
 and the trace of a matrix is represented as
  
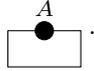
\begin{equation*}
\begin{tikzpicture}
\filldraw (0,0) circle (0.1);
  \node at (0,0.25) {$A$};
  \draw (-0.4,0)--(0.4,0)--(0.4,-0.4)--(-0.4,-0.4)--(-0.4,0);
 \end{tikzpicture} \; .
 \end{equation*}
 
Let us now consider a rank-3 tensor $A\in \mathbb{C}^D\otimes \mathbb{C}^D \otimes \mathbb{C}^d$. This is equivalent to $d$ matrices belonging to $\mathcal{M}_D$. We will denote each of these matrices as
 \begin{equation*}
A^i =
\begin{tikzpicture}
\draw (0,0)--(0,0.25);
\pic at (0,0) {tensor};
 \node at (0,0.4) {$i$};
  \node at (0,-0.25) {$A$};
  \draw (-0.4,0)--(0.4,0);
 \end{tikzpicture}
 \equiv
\begin{tikzpicture}
\filldraw (0,0.4) circle (0.08);
\draw (0,0)--(0,0.4);
\filldraw (0,0) circle (0.1);
 \node at (0.1,0.25) {$i$};
  \node at (0,-0.25) {$A$};
  \draw (-0.4,0)--(0.4,0);
 \end{tikzpicture},
 \end{equation*}
where the label above a leg is meant to fix the index to that label. The first example of a tensor network state is called Matrix Product State (MPS) \cite{Fannes92,MPSrep} and it defines a state placed on a unidimensional lattice. An MPS with periodic boundary condition (the system is placed on a ring) and constructed with local tensors independent of the site, i.e. translationally invariant, is written as follows

\begin{equation}\label{eq:MPS-TI}
\ket{\psi_A}= \sum^{d}_{i_1,\dots,i_N=1}\tr[A^{i_1}A^{i_2}\cdots A^{i_N}]\ket{i_1,\cdots,i_N} \to
 \begin{tikzpicture}
    \pic at (0,0) {tensor};
    \pic at (0.5,0) {tensor};
    \draw (-0.2,0)--(0.7,0);
     \draw (2.2,0)--(1.3,0);
    \draw (-0.2,-0.15)--(2.2,-0.15);
       \draw (2.2,0)--(2.2,-0.15);
       \draw (-0.2,0)--(-0.2,-0.15);
    \node at (1,0) {${\cdots}$};
    \pic at (1.5,0) {tensor};
    \pic at (2,0) {tensor};
    \draw (2,0)--(1.3,0);
    \end{tikzpicture} \; .
\end{equation}

Let  $D$ be the maximum rank of the virtual indices (those that get contracted) which is called bond dimension. Then, the state is specified by $ND^2d$ parameters instead of the previous exponential dependence ($d^N$) on the number of subsystems. \\

One key aspect here is how $D$ depends on $N$, since any state can be written as a tensor network with a bond dimension that grows exponentially with the number of particles. In fact, to obtain a tensor network description of any one-dimensional state successive Schmidt  Decompositions (SD) can be done \cite{Vidal}. Performing a SD between the first subsystem and the rest of the chain we obtain 
$$\ket{\psi}=\sum_{\alpha=1}^d \lambda^{[1]}_{\alpha} \ket{\alpha}^{[1]}\ket{\alpha}^{[2,\dots, N]}=\sum_{i_1=1}^d \sum_{\alpha=1}^d A^{[1]}_{i_1,\alpha}\lambda^{[1]}_{\alpha}\ket{i_1} \ket{\alpha}^{[2,\dots, N]},$$
where $A^{[1]}_{i_1,\alpha}= \langle i_1| \alpha\rangle^{[1]}$. The SD of the first two subsystems with the rest of the chain can be written as follows:
\begin{equation}\label{psiSD}\ket{\psi}=\sum_{\beta=1}^{d^2} \lambda^{[2]}_{\beta} \ket{\beta}^{[1,2]}\ket{\beta}^{[3,\dots, N]}=\sum_{i_1=1}^d \sum_{\alpha=1}^d \sum_{\beta=1}^{d^2}  A^{[1]}_{i_1,\alpha} A^{[2]}_{i_2,\alpha,\beta} \lambda^{[2]}_{\beta}\ket{i_1} \ket{i_2} \ket{\beta}^{[3,\dots, N]},
\end{equation}
where we have introduced the resolution of the identity when needed and $A^{[2]}_{i_2,\alpha,\beta}= (\bra{ \alpha}^{[1]}\bra{ i_2}) \ket{\beta}^{[1,2]}$. In this way we obtain the expression
\begin{equation}\label{MPS-OBC}
\ket{\psi}= \sum^{d}_{i_1,\dots,i_N=1}A^{[1]}_{i_1}A^{[2]}_{i_2}\cdots A^{[N]}_{i_N} \ket{i_1,\cdots,i_N} ,
\end{equation}
in which the bond dimension grows in the worst case to $d^{N/2}$ in the middle of the chain. 

Note that, even if the state $\ket{\psi}$ is translationally invariant, the description obtained in this way does not reflect this fact. In particular, it is not of the form \eqref{eq:MPS-TI}. This can be fixed, but in some cases at the price of growing the bond dimension with the size of the system \cite{qWielandt}.

The successive SD are graphically represented as:
\begin{equation*}
\begin{tikzpicture}[baseline=-1mm]
\filldraw[rounded corners=.05cm] (-1,-0.05) rectangle (1, 0.05);
  \draw (0.5,0.2)--(0.5,0);
    \draw (0.9,0.2)--(0.9,0);
      \node at (0,0.15) {${\cdots}$};
     \draw (-0.5,0.2)--(-0.5,0);
    \draw (-0.9,0.2)--(-0.9,0);
 \end{tikzpicture}
 \to 
 \begin{tikzpicture}[baseline=-1mm]
    \pic at (-0.9,0) {tensor};
    \filldraw[rounded corners=.05cm] (-0.6,-0.05) rectangle (1, 0.05);
    \draw (0.5,0.2)--(0.5,0);
    \draw (0.9,0.2)--(0.9,0);
      \node at (0,0.15) {${\cdots}$};
     \draw (-0.5,0.2)--(-0.5,0);
     \draw (-0.9,0)--(0,0);
        \end{tikzpicture}
     \to 
 \begin{tikzpicture}[baseline=-1mm]
    \pic at (-0.9,0) {tensor};
        \pic at (-0.5,0) {tensor};
    \filldraw[rounded corners=.05cm] (-0.3,-0.05) rectangle (1, 0.05);
     \draw (-0.2,0.2)--(-0.2,0);    
       \node at (0.2,0.15) {${\cdots}$};
    \draw (0.6,0.2)--(0.6,0);
    \draw (0.9,0.2)--(0.9,0);
     \draw (-0.9,0)--(0,0);
        \end{tikzpicture}
        \to
        \begin{tikzpicture}[baseline=-1mm]
            \pic at (0,0) {tensor};
    \pic at (0.5,0) {tensor};
    \draw (0,0)--(0.7,0);
    \node at (1,0) {${\cdots}$};
    \pic at (1.5,0) {tensor};
    \pic at (2,0) {tensor};
    \draw (2,0)--(1.3,0);
    \end{tikzpicture}.
\end{equation*}
But suppose that the matrices in \eqref{MPS-OBC} have size upper bounded by $D$. Then $\ket{\psi}$ satisfies the area law for any bipartition in right and left. Indeed, in that case the SD $\ket{\psi}=\sum_{\alpha=1}^D \lambda_{\alpha} \ket{\alpha}^{[R]}\ket{\alpha}^{[L]}$ has only $D$ terms, and hence the entanglement entropy $S_R(\ket{\psi})=-\sum_{\alpha=1}^D\lambda_\alpha^2\log \lambda_\alpha^2 \le \log D$ scales as the boundary of the bipartition, which in 1D is just a constant. 

The MPS-analogous tensor network states in two dimensional square lattices are the so-called PEPS \cite{Verstraete04}. A PEPS is defined by a set of rank-5 tensors $\mathcal{A}^{[v]}\in \C^d\otimes (\C^D)^{\otimes 4}$ where each $[v]$ denotes a vertex and it is represented as follows
\begin{equation*}
\ket{\Psi_\mathcal{A}}= 
 \begin{tikzpicture}
        \pic at (0,0,0.7) {3dpeps};
        \pic at (0,0,1.4) {3dpeps};   
          \pic at (0,0,0) {3dpeps};
      \pic at (0.5,0,0) {3dpeps};
      \pic at (0.5,0,0.7) {3dpeps};
      \pic at (0.5,0,1.4) {3dpeps};
      \pic at (1,0,0) {3dpeps};
        \pic at (1,0,0.7) {3dpeps};
        \pic at (1,0,1.4) {3dpeps};
         \pic at (1.5,0,0) {3dpeps};
        \pic at (1.5,0,0.7) {3dpeps};
        \pic at (1.5,0,1.4) {3dpeps};
    \end{tikzpicture},
\end{equation*}
where we will assume periodic boundary conditions, i.e. a torus, but we will not draw it.

It is not difficult to see that, for a fixed bond dimension $D$, PEPS also fulfill the area law. Hence, one can modify slightly the Area Law Conjecture and conjecture the existence of MPS or PEPS with small bond dimension that approximate well the ground state of any locally interacting gapped Hamiltonian. This can be seen as the {\it practical} version of the Area Law Conjecture since it comes with a concrete parametrization of the set of (approximate) ground states. Indeed, many algorithms (including the ubiquitous DMRG algorithm of S. White \cite{White1,White2}) aiming to solve locally interacting Hamiltonians implement different types of optimization procedures to find the MPS or PEPS that minimizes the energy. They turn out to work very well in practice (see \cite{Schollwock1,Schollwock2,Verstraete08,Ran17,Orus18} for reviews on that), supporting the validity of this modified/practical Area Law Conjecture.

At the level of mathematical proofs, it has been proven that MPS approximate well any ground state of a locally interacting gapped Hamiltonian in 1D \cite{Hastings07},\cite{Arad2013}.  Also any MPS is the (essentially unique) ground state of a locally interacting gapped Hamiltonian. So one can claim that the set of MPS {\it coincides} with the set of GS of gapped locally interacting Hamiltonians and hence gives an efficient parametrization of it. This makes MPS  the appropriate mathematical framework to prove statements about 1D systems. In 2D dimensions, despite some promising results along the same lines \cite{Hastings1,Andras}, the full picture is far from being completed. 

It is precisely the aim of this manuscript to formally state all these questions. We will separate the questions in three main topics, each of them presented in a separate section. The first one deals with questions related to the correspondence between PEPS and ground states. The second section deals with the use of PEPS to prove rigorous results in condensed-matter problems. The last section collects some open questions about PEPS that appear in different fields.

\

Let us finish this section commenting briefly on the graphical description of operators that act on the Hilbert space. For example consider an operator acting only in one site:
\begin{equation*}
\mathcal{O} \equiv
\begin{tikzpicture}
\draw (0,-0.2)--(0,0.2);
 \filldraw[draw=black, fill=red] (0,0) circle (0.06);
 \end{tikzpicture}
 \longrightarrow  \mathcal{O}_{[2]}\ket{\psi} \equiv
        \begin{tikzpicture}[baseline=-1mm]
            \pic at (0,0) {tensor};
    \pic at (0.5,0) {tensor};
      \draw (0.5,0)--(0.5,0.3);
     \filldraw[draw=black, fill=red] (0.5,0.15) circle (0.06);
    \draw (0,0)--(0.7,0);
    \node at (1,0) {${\cdots}$};
    \pic at (1.5,0) {tensor};
    \pic at (2,0) {tensor};
    \draw (2,0)--(1.3,0);
    \end{tikzpicture},
\end{equation*}
and then the expectation value is represented as:
\begin{equation}\label{ExpValue}
\langle \mathcal{O}_{[2]} \rangle \equiv
        \begin{tikzpicture}
            \pic at (0,0) {tensor};
             \pic at (0,0.3) {tensord};
    \pic at (0.5,0) {tensor};
        \pic at (0.5,0.3) {tensord};
     \filldraw[draw=black, fill=red] (0.5,0.15) circle (0.06);
    \draw (0,0)--(0.7,0);
     \draw (0,0.3)--(0.7,0.3);
    \node at (1,0) {${\cdots}$};
     \node at (1,0.3) {${\cdots}$};
    \pic at (1.5,0) {tensor};
    \pic at (1.5,0.3) {tensord};
    \pic at (2,0) {tensor};
     \pic at (2,0.3) {tensord};
    \draw (2,0)--(1.3,0);
     \draw (2,0.3)--(1.3,0.3);
    \end{tikzpicture}
    = \begin{tikzpicture}
             \pic at (-0.3,0) {tensor};
             \pic at (-0.3,-0.2) {tensord};
            \pic at (0,0) {tensor};
             \pic at (0,-0.2) {tensord};
                  \filldraw[draw=black, fill=red] (0,0.11) circle (0.05);
                 \draw (-0.3,0)--(0.15,0);
                 \draw (-0.3,-0.2)--(0.15,-0.2);
                                   \draw[preaction={draw, line width=1pt, white}][line width=0.5pt] (0,0.19) to [out=45, in=-45] (0,-0.39);
                                                     \draw[preaction={draw, line width=1pt, white}][line width=0.5pt] (-0.3,0.19) to [out=45, in=-45] (-0.3,-0.39);
                 \node at (0.35,0) {${\cdots}$};
                 \node at (0.35,-0.2) {${\cdots}$};
    \pic at (0.7,0) {tensor};
    \pic at (0.7,-0.2) {tensord};
                         \draw (0.55,0)--(1,0);
                     \draw (0.55,-0.2)--(1,-0.2);
                    \draw[preaction={draw, line width=1pt, white}][line width=0.5pt] (0.7,0.19) to [out=45, in=-45] (0.7,-0.39);
    \pic at (1,0) {tensor};
    \pic at (1,-0.2) {tensord};
                \draw[preaction={draw, line width=1pt, white}][line width=0.5pt] (1,0.19) to [out=45, in=-45] (1,-0.39);
    \end{tikzpicture}
    \equiv \tr[\mathcal{O}_{[2]} \ket{\psi}\bra{\psi}].
\end{equation}
In general for an operator acting on $N$ sites the representation is the following:
\begin{equation*}
\begin{tikzpicture}[baseline=-1mm]
\filldraw[rounded corners=.05cm] (-1,-0.05) rectangle (1, 0.05);
  \draw (0.5,0.2)--(0.5,-0.2);
    \draw (0.9,0.2)--(0.9,-0.2);
      \node at (0,0.15) {${\cdots}$};
            \node at (0,-0.15) {${\cdots}$};
     \draw (-0.5,0.2)--(-0.5,-0.2);
    \draw (-0.9,0.2)--(-0.9,-0.2);
 \end{tikzpicture}.
 \end{equation*}
But one can consider operators coming from a tensor network, that is, Matrix Product Operators (MPO):
\begin{equation*}
        \begin{tikzpicture}[baseline=-1mm]
            \pic at (0,0) {tensorud};
    \pic at (0.5,0) {tensorud};
    \draw (0,0)--(0.7,0);
    \node at (1,0) {${\cdots}$};
    \pic at (1.5,0) {tensorud};
    \pic at (2,0) {tensorud};
    \draw (2,0)--(1.3,0);
    \end{tikzpicture},
\end{equation*}
where the local tensor are matrices depending on two virtual indices.  

\section{Are PEPS and GS of local gapped Hamiltonians the same set?}

As commented in the introduction, one of the key features of PEPS is that they are conjectured to {\it correspond}  to the set of ground states of gapped and locally interacting Hamiltonians (modified Area Law Conjecture). This is motivated by the fact that this is the situation for the one dimensional case with MPS. This correspondence can be divided in two statements:

\begin{enumerate}
\item[1] Ground states of gapped locally interacting Hamiltonians can be well approximated by PEPS with a \emph{small} bond dimension (i.e. GS $\subset$ PEPS).
\item[2] PEPS are exact ground states of (gapped) locally interacting Hamiltonians (i.e. PEPS $\subset$ GS).
\end{enumerate}

Some comments are in order: 

As shown in \cite{Ge}, there are examples of states in 2D that fulfill an area law and, however, are not ground states of local Hamiltonians (nor well approximated by PEPS). In this sense, the set of area-law states is too big to capture the desired set of ground states and it is precisely the family of PEPS the one that seems to capture better such set.

The gap in statement 2 cannot be always guaranteed, as there are examples of PEPS that cannot be ground states of any gapped Hamiltonian \cite{Ising-PEPS}. This will be commented in detail in subsection \ref{sec:spectral-gap} below.

\

In the following we will pose the main open questions concerning points 1 and 2, together with the state of the art for both of them. 
\subsection{Are all PEPS the GS of a local gapped Hamiltonian?}

Every PEPS is the GS of a locally interacting Hamiltonian, called {\it parent Hamiltonian}. We revise the construction here. 

Let us start with a translationally invariant PEPS on a $L\times L$ torus defined by a tensor $\mathcal{A}$. (The construction can be easily generalized to other sizes and geometries, and also to the absence of translation invariance, but let us stick to this case for the sake of simplicity.) Take a square region $\mathcal{R}$ of size $n\times n$ in the torus and define for that region the linear map 
$$\Gamma_\mathcal{R}: (\C^D)^{\otimes |\partial \mathcal{R}|}\rightarrow (\C^d)^{\otimes |\mathcal{R}|}$$
which maps, using the tensors $\mathcal{A}$ in $\mathcal{R}$, boundary conditions living in the virtual space to vectors in the physical Hilbert space of the region $\mathcal{R}$. With the graphical notation of section \ref{sec:TNS}:

\begin{equation*}
X\equiv
 \begin{tikzpicture}
     \begin{scope}[canvas is zx plane at y=0]
      \draw[rounded corners=.05cm, very thick] (-0.53,-0.4) rectangle (1.9,1.9);
    \end{scope}
        \draw (-0.4,0,0.7) -- (-0.2,0,0.7);
       \draw (-0.4,0,1.4) -- (-0.2,0,1.4);   
        \draw (-0.4,0,0) -- (-0.2,0,0);   
          \draw (0,0,-0.5)--(0,0,-0.2);
            \draw (0.5,0,-0.5)--(0.5,0,-0.2);
            \draw (1,0,-0.5)--(1,0,-0.2);
            \draw (1.5,0,-0.5)--(1.5,0,-0.2);
          \draw (0,0,1.6)--(0,0,1.9);
          \draw (0.5,0,1.6)--(0.5,0,1.9);
          \draw (1,0,1.6)--(1,0,1.9);
           \draw (1.5,0,1.6)--(1.5,0,1.9);
       \draw (1.9,0,0.7) -- (1.7,0,0.7);
        \draw (1.9,0,0) -- (1.7,0,0);
         \draw (1.9,0,1.4) -- (1.7,0,1.4);
    \end{tikzpicture}
\longrightarrow
 \begin{tikzpicture}
     \begin{scope}[canvas is zx plane at y=0]
      \draw[rounded corners=.05cm, very thick] (-0.53,-0.4) rectangle (1.9,1.9);
    \end{scope}
        \pic at (0,0,0.7) {3dpeps};
        \pic at (0,0,1.4) {3dpeps};   
          \pic at (0,0,0) {3dpeps};
      \pic at (0.5,0,0) {3dpeps};
      \pic at (0.5,0,0.7) {3dpeps};
      \pic at (0.5,0,1.4) {3dpeps};
      \pic at (1,0,0) {3dpeps};
        \pic at (1,0,0.7) {3dpeps};
        \pic at (1,0,1.4) {3dpeps};
         \pic at (1.5,0,0) {3dpeps};
        \pic at (1.5,0,0.7) {3dpeps};
        \pic at (1.5,0,1.4) {3dpeps};
    \end{tikzpicture}
    \equiv \Gamma_R(X).
\end{equation*}

Let us denote $\mathcal{G}_\mathcal{R}={\rm Im}(\Gamma_\mathcal{R})$ and define the interaction term $h_\mathcal{R}$ as the orthogonal projection onto $\mathcal{G}_\mathcal{R}^\perp$ that is, 
$\ker(h_\mathcal{R})=\mathcal{G}_\mathcal{R}$. The Hamiltonian is then defined by translating $h_R$, $H=\sum_\tau \tau(h_\mathcal{R})$, where the sum runs over all possible translations in the torus. It is clear that the given PEPS is a ground state of $H$ and that $H$ is frustration free, meaning that the ground state of $H$ minimizes the energy of the local term $h_\mathcal{R}$; $h_\mathcal{R}\ket{\Psi_A}=0$. The basic open question here is the following 

\begin{question}
Which are the minimal requirements on $\mathcal{A}$ and the minimal size of $\mathcal{R}$ under which one can guarantee that the given PEPS is the unique ground state of $H$ and in addition $H$ is gapped? 
\end{question}

This question turns out to be very difficult, specially beyond 1D systems. Let us now go slowly  through the known results and divide this question into more specific ones. For that, we introduce the key concepts of {\it normal} and {\it injective} tensors which endow $\mathcal{A}$ with some special properties.

\begin{definition}\label{def:injectivity}
A tensor $\mathcal{A}$ is called injective if, viewed as a linear map from the virtual indices to the physical space, it is an injective map. With the above notation, this is just saying that $\Gamma_\mathcal{R}$ is an injective map for the region $\mathcal{R}$ of size $1\times 1$. 
\end{definition}

\begin{definition}\label{def:normality} A tensor $\mathcal{A}$ is {\it normal} if there exists $n\in\N$ so that $\Gamma_\mathcal{R}$ is an injective map for the square region $\mathcal{R}$ of size $n\times n$. In that case, the smallest such $n$ is the {\it injectivity index} of $\mathcal{A}$ and we denote it by $i(\mathcal{A})$.
\end{definition}

It is known \cite{Fannes92,reviewPEPS} that, given a normal tensor $\mathcal{A}$ with injectivity index $i(\mathcal{A})$, by taking $\mathcal{R}$ as the square region of size $i(\mathcal{A})+1$, the parent Hamiltonian associated to $\mathcal{R}$ with the above construction has the PEPS $\ket{\Psi_\mathcal{A}}$ as the unique ground state with zero energy $H\ket{\Psi_\mathcal{A}}=0$. 

Therefore, the bounds on the injectivity index correspond to the bounds on the interaction length of the parent Hamiltonian.  To comment on such bounds we will start with the case of 1D. There, in order to briefly illustrate about the techniques used so far, we will make a small detour and talk about a classic inequality of Wielandt in the context of stochastic matrices.

\subsubsection{Wielandt inequalities}

In 1950 \cite{Wielandt}, Wielandt proved that the index of primitivity of a primitive stochastic matrix $A\in \mathcal{M}_{D\times D}$ must be less or equal than $D^2-2D+2$, and that this bound is optimal. 

Let us recall that an stochastic matrix $A= (A_{i,j})_{i,j}\in \mathcal{M}_{D\times D}$ is a matrix with $A_{i,j}\ge 0$ for all $i,j$ and $\sum_{i} A_{i,j}=1$ for all $j$. This implies that if $p=(p_i)_i$ is a probability distribution ($p_i\ge 0$ and $\sum_i p_i=1$), the same holds for $Ap=(\sum_j A_{i,j}p_j)_i$. In this sense, $A$ models a noisy memoryless communication channel acting on an alphabet of size $D$ -- the basic object in Shannon's information theory.  

A stochastic matrix $A$ is called primitive if there exists $n\in \N$ such that $(A^n)_{i,j}>0$ for all $i,j$. The minimum of such $n$ is called the index of primitivity of $A$. 

The range of applications of Wielandt's inequality is wide: Markov chains \cite{Seneta}, graph theory and number theory \cite{Alfonsin}, or numerical analysis \cite{Varga} to name a few. 

In quantum information theory, the object that models a memoryless noisy channel is a trace-preserving completely positive linear map (also called quantum channel) $T:\mathcal{M}_{D\times D} \rightarrow \mathcal{M}_{D\times D}$ \cite{NielsenChuang}. The quantum channel  $T$, by means of its Kraus decomposition, is nothing but a map of the form $T(X)=\sum_{i=1}^d A_i X A_i^{\dagger}$, where the {\it Kraus operators} $A_i$ are $D\times D$ matrices fulfilling $\sum_i A_i^{\dagger} A_i=\1$ (this is precisely the trace preserving condition) and $A^{\dagger}$ denotes the adjoint matrix of $A$.

Note that quantum channels include stochastic matrices as particular cases. Given a stochastic matrix $A=(a_{i,j})$, the quantum channel  $T_A$ with Kraus operators $\sqrt{a_{i,j}} \ket{i}\bra{j}$ has the following property: given a probability vector $p$, if we consider the diagonal matrix $\rho={\rm diag} (p)=\sum_i p_i \ket{i}\bra{i}$ then, $T_A(\rho)$ is exactly ${\rm diag} (Ap)$.  That is, the quantum channel $T_A$ restricted to the diagonal matrices is exactly the stochastic matrix $A$.

The following definition is the natural quantum (non-commutative) analogue of the notion of primitivity for a stochastic matrix \cite{qWielandt}.

\begin{definition}
A quantum channel is called primitive if there exists an $n\in \N$ so that $T^n(\rho)$ is full rank for all positive semi-definite input $\rho$.   The minimum of such $n$ is called the primitivity index $p(T)$. 
\end{definition}

Note that given a stochastic matrix $A$, the associated quantum channel $T_A$ is primitive if and only if $A$ is primitive. Moreover, the corresponding primitivity indices coincide. There is an equivalent notion of primitive quantum channel, related to the classical Perron-Frobenius-like characterization of primitivity for the stochastic case \cite{qWielandt}:

\begin{proposition}
A quantum channel $T$ is primitive if and only if $T$ has a unique non-degenerate eigenvalue $\lambda$ with $|\lambda|=1$ and the corresponding eigenvector (which is necessarily semi-definite positive) is full rank. 
\end{proposition}

A natural question arises:

\begin{question}
Which is the optimal upper bound for the primitivity index of a primitive quantum channel $T$ acting on $\mathcal{M}_{D\times D}$?
\end{question}

In \cite{qWielandt} it is shown that $p(T)\le (D^2-d+1)D^2$. This result has been recently improved \cite{Rahaman} to $p(T)\le 2(D-1)^2$. The order $O(D^2)$ is optimal just by invoking the optimality of the classical Wielandt inequality. However, the exact optimal bound is still unknown

As shown in \cite{qWielandt}, this type of bounds gives universal thresholds for the behavior in time of the {\it zero-error classical capacity of a quantum channel}, denoted by $C_0(T)$, defined as the optimal rate (measured in number of bits per use of the channel) at which a quantum channel can transmit classical information without errors \cite{Leung}. The following dichotomy result can be shown \cite{qWielandt}: \\
\begin{proposition}
Let $T$ be a quantum channel with a full-rank fixed point. Then, either $C_0(T^n)\ge1$ for all $n\in \N$ or $C_0(T^{p(T)})=0$,
\end{proposition}

\subsubsection{Index of injectivity of a MPS}

Let us now connect the previous discussion with the injectivity index of an MPS, as defined in Definition \ref{def:normality}.

We recall that a translationally invariant MPS is given by a rank-3 tensor $A$, which is nothing but a set of matrices $A_i\in \mathcal{M}_{D\times D}$, $i=1,\ldots, d$, and hence it naturally defines a completely positive linear map $\mathcal{E}_{{A}}(X)= \sum_i A_i XA_i^\dagger$. Such map is usually called the transfer operator associated to the MPS. Using the transformation $A_i\mapsto YA_iY^{-1}$ that leaves invariant the MPS $\ket{\psi_A}$, one can assume w.l.o.g that the transfer operator $\mathcal{E}_{A}$ is trace-preserving and hence a quantum channel (see e.g. \cite{MPDO} for details).

It is easy to see \cite{qWielandt} that the MPS is injective if and only if its associated transfer operator is primitive. In the normal case, the injectivity index of an MPS is an upper bound to the index of primitivity of its associated transfer operator, i.e. $i(A)\ge p(\mathcal{E}_{A})$. This finally brings us to the following key question: 

\begin{question} \label{question:injectivity-length-MPS}
Which is the optimal upper bound for the injectivity index of a normal MPS in terms of its bond dimension?
\end{question}

In \cite{qWielandt} it is shown that if $A$ is normal then $i({A})\le (D^2-d+1)D^2$. This result has been recently improved \cite{Michalek} to $p(T)\le 2D^2(6+\log_2(D))$. Up to a   logarithmic factor, the order $O(D^2\log(D))$ is optimal just by invoking the optimality of the classical Wielandt inequality. As before, the exact optimal bound is still unknown.\\

\subsubsection{Index of injectivity of a PEPS}

Motivated by the connection between the injectivity index and the interaction length of the parent Hamiltonian, one may ask the analogue of Question \ref{question:injectivity-length-MPS} in 2D.

\begin{question} \label{question:injectivity-PEPS1}
Which is the optimal upper bound for the injectivity index of a normal  PEPS in terms of its bond dimension?
\end{question}

As opposed to the 1D case, the only known result, proven recently in \cite{Michalek-2D} is the existence of a function of the bond dimension $f(D)$ that bounds $i(\mathcal{A})$ for every PEPS with bond dimension $D$. Unfortunately, in principle such function could be uncomputable.

Indeed, checking normality becomes undecidable if one generalizes the notion of normal PEPS as those tensors $\mathcal{A}$ with the following properties: 
\begin{enumerate}
\item There exists an orthogonal projector $P:\mathbb{C}^D\to \mathbb{C}^D$ so that the tensor $\mathcal{B}=(\1_d \otimes P^{\otimes 4}) \mathcal{A}$ is normal and
\item the PEPS associated to $\mathcal{A}$ and $\mathcal{B}$ coincide for every system size, i.e. $\ket{\Psi_{\mathcal{A}}}= \ket{\Psi_{\mathcal{B}}}$.
\end{enumerate}
This can be proven easily with the techniques in \cite{uncomputable-PEPS}. Therefore a weaker version of  Question \ref{question:injectivity-PEPS1} should be considered:

\begin{question} \label{question:injectivity-PEPS2}
Give an explicit (computable and if possible polynomial) function $f(D)$ which is an upper bound for the injectivity index of a normal PEPS.
\end{question}

\subsubsection{Spectral gap in PEPS} \label{sec:spectral-gap}

Let us finish this subsection tackling the problem of the spectral gap of the parent Hamiltonian. It is proven in \cite{Fannes92} (see \cite{Katroyano-boundary} for an alternative proof) that the parent Hamiltonian of a normal MPS is always gapped.  Unfortunately, this is not the case for 2D in PEPS, as it is shown in \cite{Ising-PEPS} by constructing an explicit counterexample.

In fact, for general PEPS the existence of gap in the parent Hamiltonian is undecidable, as shown in \cite{uncomputable-PEPS}, which highlights the complexity of the problem.  Moreover, the spectral gap of even the simplest non-trivial PEPS --the AKLT model \cite{AKLT} as the paradigmatic example-- is still open.

However, some light has been shed on checking whether a Hamiltonian is gapped or not translating the question into a problem on the boundary. For instance, in \cite{Didier}, motivated by the holographic correspondence uncovered by Li and Haldane in \cite{LiHaldane}, an exact bulk-boundary correspondence was found, constructing for every PEPS a (family of) 1D mixed states, named as boundary states. In that work it is conjectured (see also \cite{Katroyano-boundary}), based on numerical evidence that

\begin{conj} \label{gap2Dboundary1dlocal}The gap of the parent Hamiltonian of a PEPS corresponds exactly to the possibility of writing the boundary states as Gibbs states of 1D short range Hamiltonians 
$$\rho=e^{-\beta H}, \quad \text{with } H=\sum_{i,j} h_{i,j},\quad \|h_{i,j}\|\le Je^{-\alpha |i-j|},$$
where $h_{i,j}$ acts non-trivially only on spins $i$ and $j$. 
\end{conj}

The boundary states are simply the semi-definite operators defined on $(\C^D)^{\otimes |\partial \mathcal{R}|}$ obtained in the boundary of a region when tracing out the bulk as shown in the figure

\begin{equation*}
\rho_\mathcal{R}=
 \begin{tikzpicture}
             \pic at (-0.3,0) {tensor};
             \pic at (-0.3,-0.2) {tensord};
            \pic at (0,0) {tensor};
             \pic at (0,-0.2) {tensord};
                 \draw (-0.3,0)--(0.15,0);
                 \draw (-0.3,-0.2)--(0.15,-0.2);
                 \node at (0.35,0) {${\cdots}$};
                 \node at (0.35,-0.2) {${\cdots}$};
    \pic at (0.7,0) {tensor};
    \pic at (0.7,-0.2) {tensord};
                     \draw (0.55,0)--(1.15,0);
                     \draw (0.55,-0.2)--(1.15,-0.2);
                         \node at (0.85,0.1) {$\alpha$};
                         \node at (0.85,-0.35) {$\beta$};
    \pic at (1,0) {tensor};
    \pic at (1,-0.2) {tensord};
                \draw[preaction={draw, line width=1pt, white}][line width=0.5pt] (1,0.19) to [out=45, in=-45] (1,-0.39);
    \node at (1.35,0) {${\cdots}$};
                 \node at (1.35,-0.2) {${\cdots}$};
                  \pic at (1.7,0) {tensor};
             \pic at (1.7,-0.2) {tensord};
                               \pic at (2,0) {tensor};
             \pic at (2,-0.2) {tensord};
             \draw (2,0)--(1.55,0);
                 \draw (2,-0.2)--(1.55,-0.2);
            \draw[preaction={draw, line width=1pt, white}][line width=0.5pt] (1.7,0.19) to [out=45, in=-45] (1.7,-0.39);
                        \draw[preaction={draw, line width=1pt, white}][line width=0.5pt] (2,0.19) to [out=45, in=-45] (2,-0.39);
    \end{tikzpicture}
    = \sum_{\alpha,\beta} (\ket{\Psi^{[\mathcal{R}]}_\mathcal{A}})_\alpha  (\sigma_{\mathcal{R}^c})_{\alpha,\beta}(\bra{\Psi^{[\mathcal{R}]}_\mathcal{A}})_\beta
\end{equation*}
As in any holographic correspondence, one is interested in creating a dictionary that maps bulk properties to boundary properties. The reason that such dictionary is expected in PEPS comes from the way in which expectation values are computed (see Eq.(\ref{ExpValue}) in Section \ref{sec:TNS} ): the boundary states are exactly the operators that mediate at the virtual level the correlations present at the physical level. Then,
 
\begin{question}
Is Conjecture \ref{gap2Dboundary1dlocal} true?
\end{question}

An important step in this direction was given in \cite{Katroyano-boundary}, proving one of the implications for the case of a faster than exponential decay in $\|h_{i,j}\|$.

\subsection{Can any GS of a local gapped Hamiltonian be represented as a PEPS?}

One of the main features of PEPS, and the one that makes them a relevant ansatz in the classical simulation of quantum systems, is the conjectured fact that PEPS approximate well ground states of locally interacting gapped Hamiltonians. To formalize this, we consider a gapped, translationally invariant Hamiltonian on an $L\times L$ torus given by a finite range interaction $h$, $H=\sum_\tau \tau(h)$. We will assume a unique ground state denoted by  $\ket{\Psi_{\rm GS}}$.

There are two types of relevant approximations, global and local, depending on whether one is interested in approximating an extensive or an intensive quantity in the ground state. 

In the global approximation problem, the aim is to find a function $f(L)$ such that one can guarantee the existence of a (non-necessarily translationally invariant) PEPS $\ket{\Psi_{\rm PEPS}}$ with bond dimension $D\le f(L)$ so that in the Hilbert norm
$$\|\ket{\Psi_{\rm PEPS}}-\ket{\Psi_{\rm GS}}\|_2\le \frac{1}{{\rm poly}(L)} \; .$$

For the local approximation problem, the goal is to find a function, if it exists, $g(\epsilon)$, so that one can guarantee the existence of a translationally invariant PEPS $\ket{\Psi_\mathcal{A}}$ given by a tensor $\mathcal{A}$, with bond dimension $D\le g(\epsilon)$, so that in trace-class norm, 
$$\lim_{L\rightarrow \infty} \|\rho^{[L]}_{\mathcal{R}, GS}-\rho^{[L]}_{\mathcal{R}, \mathcal{A}}\|_1\le \epsilon\; ,$$
where $\rho^{[L]}_{\mathcal{R}, GS}$ is the reduced density matrix of the region $\mathcal{R}$ associated to $\ket{\Psi_{\rm GS}}$ in the torus of size $L\times L$ ($\rho^{[L]}_{\mathcal{R}, \mathcal{A}}$ is defined analogously). Note that being both $\ket{\Psi_\mathcal{A}}$ and $\ket{\Psi_{\rm GS}}$ translationally invariant, the exact position of region $\mathcal{R}$ in the torus is irrelevant.  This type of approximation guarantees that in the thermodynamic limit, compactly supported observables can be well approximated by translationally invariant PEPS (with finite bond dimension).

Both the global and local approximation problems have a positive satisfactory solution in 1D, with the current best bounds being  
\begin{align}\label{eq:bounds-approx}
f(L))&= e^{O(\log^{3/4}L)} \\
g(\epsilon)&= e^{O(\log^{3/4}\frac{1}{\epsilon})} \nonumber \; .
\end{align}
 proven in \cite{Arad2013} and \cite{Huang} respectively .

Both results come from refined versions of the so-called detectability lemma \cite{Aharonov,Anshu}. For simplicity, we will state it in 1D for nearest-neighbor interactions but a similar result holds in any dimension for finite range interactions. 

\begin{lemma}[Detectability Lemma in 1D]\label{lemma:DL}
Let $P$ be an orthogonal  projector on $\C^d\otimes \C^d$ and $Q=\1-P$ its orthogonal complement. Denote by $P_i$ the projector $P$ acting on sites $i,i+1$ of a chain of $L$ spins with periodic boundary conditions. Let $H=\sum_{i=1}^L P_i$ be a frustration free Hamiltonian and let 
$DL(H)$ be the operator
$$DL(H)=\left(\bigotimes_{i\text{ even}} Q_i\right)\left(\bigotimes_{i\text{ odd}} Q_i\right)$$ 

Then $$\left\| \ket{\Psi_{\rm GS}}\bra{\Psi_{\rm GS}} - DL(H)^\ell\right\|_\infty\le \left(\frac{1}{\sqrt{\frac{\Delta}{4}+1}} \right)^\ell= e^{-\alpha\ell},$$
where $\Delta$ is the spectral gap of $H$ (and $\alpha=\frac{1}{2}\log(\frac{\Delta}{4}+1)$).
\end{lemma}

To get an intuition of its application, let us briefly show how to use Lemma \ref{lemma:DL} to show approximation in operator norm of the ground state projector of $H$ by a MPO. Each $Q_i$ in $DL(H)$ is a two-body operator so both operators can be represented graphically as:

\begin{equation*}
Q_i= 
\begin{tikzpicture}
\pic at (0,0) {O2};
\end{tikzpicture}
\Rightarrow DL(H)=
\begin{tikzpicture}
 \node at (-0.6,0.15) {${\cdots}$};
  \pic at (0,0) {O2};
  \pic at (0.6,0.3) {O2};  
   \pic at (1.2,0) {O2};
     \pic at (1.8,0.3) {O2};  
        \pic at (2.4,0) {O2};
     \pic at (3,0.3) {O2}; 
      \node at (3.6,0.15) {${\cdots}$};
      \end{tikzpicture}.
 \end{equation*}
Then, by doing a SVD decomposition in each $Q_i$; 
 \begin{equation*}
\begin{tikzpicture}
\pic at (0,0) {O2};
\end{tikzpicture}
=
\begin{tikzpicture}
\draw (-0.3,0)--(0.3,0);     
  \draw (-0.3,0.2)--(-0.3,-0.2);
    \draw (0.3,0.2)--(0.3,-0.2);
     \filldraw[draw=black, fill=red] (0,0) circle (0.06);
      \filldraw[draw=black, fill=blue] (-0.35,-0.05) rectangle (-0.25,0.05);
        \filldraw[draw=black, fill=blue] (0.25,-0.05) rectangle (0.35,0.05);
             \node at (-0.47,-0.15)  {$U$};
                \node[anchor=north] at  (0,0)  {$\Sigma$};
                \node at (0.5,-0.13)  {$V^\dagger$};
 \end{tikzpicture}
 \equiv
 \begin{tikzpicture}
\pic at (0,0) {O2SVD};
 \end{tikzpicture},
 \end{equation*}
 it is easy to see that $DL(H)^{\ell}$ is an MPO with bond dimension $D\le d^{2\ell}$:
 \begin{equation*}
 \begin{tikzpicture}
  \node at (-0.3,0.075) {$\big($};
  \pic at (0,0) {O2p};
  \pic at (0.3,0.15) {O2p};  
   \pic at (0.6,0) {O2p};
     \pic at (0.9,0.15) {O2p};  
      \node at (1.2,0.075) {$\big)$};
            \node at (0.5,0.3) {${\cdot}$};
            \node at (0.5,0.35) {${\cdot}$};
              \node at (0.5,0.4) {${\cdot}$};
  \node at (-0.3,0.625) {$\big($};
  \pic at (0,0.55) {O2p};
  \pic at (0.3,0.7) {O2p};  
   \pic at (0.6,0.55) {O2p};
     \pic at (0.9,0.7) {O2p};  
      \node at (1.2,0.625) {$\big)$};
      \end{tikzpicture}
      \equiv
  \begin{tikzpicture}
  \node at (-0.45,0.085) {$\big($};
  \pic at (0,0) {O2SVD};
          \pic at (0.8,0) {O2SVD}; 
      \pic at (0.4,0.17) {O2SVD};  
     \pic at (1.2,0.17) {O2SVD}; 
      \node at (1.45,0.085) {$\big)$};
            \node at (0.6,0.3) {${\cdot}$};
            \node at (0.6,0.35) {${\cdot}$};
              \node at (0.6,0.4) {${\cdot}$};
       \node at (-0.45,0.635) {$\big($};       
  \pic at (0,0.55) {O2SVD};
          \pic at (0.8,0.55) {O2SVD}; 
      \pic at (0.4,0.72) {O2SVD};  
     \pic at (1.2,0.72) {O2SVD}; 
      \node at (1.45,0.635) {$\big)$};
      \end{tikzpicture}.
 \end{equation*}

Now, fixing $\epsilon$ and solving $\epsilon= e^{-\alpha\ell}$ (see Lemma \ref{lemma:DL}), we get $\ell= \frac{1}{\alpha}\log\frac{1}{\epsilon}$ and, by Lemma \ref{lemma:DL}, the operator $DL(H)^\ell$ approximates within $\epsilon$ the ground state projector on operator norm and has bond dimension 	
$$D\le d^{2\ell}=d^{\frac{2}{\alpha}\log\frac{1}{\epsilon}}= \left(\frac{1}{\epsilon}\right)^{\frac{\log d^2}{\alpha}}={\rm poly}\left(\frac{1}{\epsilon}\right).$$

In order to have the required approximation in trace class norm, and to get it beyond frustration free systems, more sophisticated versions of Lemma \ref{lemma:DL} are required \cite{Arad2013}, leading to the bounds of Eq.(\ref{eq:bounds-approx}).

However in 2D the analogue problems are quite open. First of all, there is no known solution of the local approximation problem. Second, the best known function associated to the global approximation problem is superpolynomial $f(L)=e^{O(\log^2 L)}$ and, moreover, it can only be guaranteed to work under extra spectral assumptions on the Hamiltonian. Specifically, under the following assumption about the absence of concentration of eigenvalues close to the ground state energy: for each $M>0$, the number of eigenstates with energy lower than $E_0+M$ grows at most polynomially with the system size $L$. 
 
 Three questions arise here which can be seen as variants of the Area Law Conjecture.

\begin{question}
Does there exist a global approximation result in 2D only under the spectral gap assumption?
\end{question}

\begin{question}
Can the function $f(L)$ be taken polynomial in $L$?
\end{question}

\begin{question}
Does there exist a local approximation result in 2D? Is this possible assuming only the spectral gap assumption? Can $g(\epsilon)$ be taken polynomial in $\frac{1}{\epsilon}$?
\end{question}

\section{PEPS as a framework to give formal proofs in cond-mat problems}

The results and questions stated in the previous section point to the informal statement that $PEPS = GS$. This opens the possibility to analyze relevant questions  for GS, that are really hard to solve in the case of arbitrary systems, using the framework of PEPS where rigorous mathematical proofs can be found.

An illustrative example is the study of 1D GS that are invariant under symmetries. In particular the question is the following: {\it in how many different ways a group can act as a symmetry in a quantum many body system?} The inequivalent ways of that action classify the so-called Symmetry Protected Topological (SPT) phases and they are defined formally as follows:

\begin{definition}
Consider two gapped locally interacting Hamiltonians $H_0=\sum_\tau \tau(h_0)$ and $H_1=\sum_\tau \tau(h_1)$ on a ring $\Lambda$,  supported on local Hilbert spaces $\mathcal{H}_0=\C^{d_0}$ and $\mathcal{H}_1=\C^{d_1}$ respectively and such that they commute with unitary representations $U_0:G\rightarrow  \mathcal{U}(d_0)$, $U_1:G\rightarrow  \mathcal{U}(d_1)$ of a group $G$ (meaning that $[H_i,U_i(g)^{\otimes |\Lambda|}]=0$ for all $g\in G$) respectively. We say that $H_0$ and $H_1$ are in the same SPT phase if there exist another local ancillary Hilbert space $\mathcal{H}_a=\C^{d_a}$ and a locally interacting Hamiltonian $H_\lambda=\sum_\tau \tau(h_\lambda)$ with local Hilbert space $\mathcal{H}=\mathcal{H}_0\oplus  \mathcal{H}_1\oplus  \mathcal{H}_a$ so that
\begin{enumerate}
\item $[0,1]\ni\lambda\mapsto h_\lambda$ is smooth (real analytic) (where $\mathcal{H}_0$ and $\mathcal{H}_1$ are embedded in the corresponding sector of $\mathcal{H}$.)
\item There exists a representation $U_a:G\mapsto \mathcal{U}( d_a)$ so that $H_\lambda$ commutes with $(U_0\oplus U_1\oplus U_a)^{\otimes |\Lambda|}$  for all $\lambda$.
\item The spectral gap of $H_\lambda$ is bounded from below by a constant $c>0$ which is independent of $\lambda$ and the system size $|\Lambda|$.
\end{enumerate}
\end{definition}

It is clear that this definition gives rise to an equivalence relation, the different equivalent classes being the different SPT phases. Then, one can rephrase the question by: \emph{how many SPT phases are there for a given group $G$?}

In order to solve this question in unique GS of local gapped Hamiltonians, one can restrict to the case of injective MPS (and their parent Hamiltonians) that are invariant under the action of a symmetry, i.e. MPS so that 
\begin{equation}\label{eq:MPS-sym}
\ket{\psi_{A}}=U(g)^{\otimes L}\ket{\psi_{A}}.
\end{equation}
It is proven in \cite{SPT} that Eq.\eqref{eq:MPS-sym} holds for all $L$ if and only if there exists a projective representation $V_g$ of $G$ acting on the virtual space $\mathcal{M}_{D\times D}$ so that

 \begin{equation} \label{fig:local-sym-MPS}
    \begin{tikzpicture}[baseline=-1mm]
      \pic at (0,0) {tensor};
      \draw (-0.25,0) -- (0.25,0);
      \draw (0,0) -- (0,0.25);
      \filldraw[draw=black, fill=purple] (0,0.15,0) circle (0.04);
      \node[anchor=east] at (0,0.25,0) {$U(g)$};
    \end{tikzpicture} =
    \begin{tikzpicture}[baseline=-1mm]
         \pic at (0,0) {tensor};
      \draw (-0.3,0) -- (0.3,0);
      \filldraw [draw=black, fill=red] (-0.2,0,0) circle (0.04);
      \filldraw [draw=black, fill=red] (0.2,0,0) circle (0.04);
      \node[anchor=north] at (-0.25,0.05) {$\myinv{V_g}$};
      \node[anchor=north] at (0.25,0) {$V_g$};
          \end{tikzpicture},
  \; \forall g\in G.
  \end{equation} 
 
From there, one can prove \cite{SPT} that 1D SPT phases in MPS are exactly given by the different non-equivalent projective representations of $G$, which is exactly the second cohomology group $H^2(G,U(1))$.

\subsection{Fundamental theorem in PEPS}

It is clear from the above argument how crucial it is to have a local (single tensor) characterization as in Eq. (\ref{fig:local-sym-MPS}) of the existence of a global symmetry \eqref{eq:MPS-sym}. In fact such local characterization is just a particular case, by fixing $g$ and defining $\mathcal{B}= U(g) \mathcal{A}$, of the following more general question for PEPS:
\begin{question}
What is the relation between two tensors $\mathcal{A}$ and $\mathcal{B}$  that define the same PEPS, i.e. $\ket{\Psi_\mathcal{A}}=\ket{\Psi_\mathcal{B}}$, on a torus $L\times L$ for all possible sizes $L$?
\end{question}

The so-called {\it Fundamental Theorem of MPS} \cite{MPDO} shows that this happens in 1D if and only if there exists an invertible matrix $Y$ so that 
 \begin{equation} 
    \begin{tikzpicture}[baseline=-1mm]
      \pic at (0,0) {tensor};
      \draw (-0.25,0) -- (0.25,0);
      \draw (0,0) -- (0,0.25);
      \node[anchor=north] at (0,0) {${B}$};
    \end{tikzpicture} =
    \begin{tikzpicture}[baseline=-1mm]
         \pic at (0,0) {tensor};
         \node[anchor=north] at (0,0) {${A}$};
      \draw (-0.3,0) -- (0.3,0);
      \filldraw [draw=black, fill=red] (-0.2,0,0) circle (0.04);
      \filldraw [draw=black, fill=red] (0.2,0,0) circle (0.04);
      \node[anchor=north] at (-0.3,0.05) {$\myinv{Y}$};
      \node[anchor=north] at (0.3,0) {$Y$};
          \end{tikzpicture}.
  \end{equation} 
The reason behind the name {\it Fundamental Theorem} stems from the fact that it is ubiquitously used in very different contexts, ranging from quantum cellular automata \cite{QCA} to the description of 2D topological orders \cite{MPDO}. Formally, one needs to bring first ${A}$ and ${B}$ to the so-called canonical form to state the Fundamental Theorem --see \cite{MPDO} for details.

For two and larger dimensions, a similar result holds true if the tensors $\mathcal{A}$ and $\mathcal{B}$ are normal \cite{Carlos,Molnar1}. Unfortunately, as opposed to the 1D case, in 2D  the restriction to normal tensors excludes all non-trivial SPT phases. This is why extending the Fundamental Theorem in 2D beyond normal tensors becomes a crucial question to solve (see \cite{Molnar2} for one such extension to the case of so-called quasi-injective PEPS).

On the opposite direction, it is shown in \cite{uncomputable-PEPS} that it is undecidable to know whether two general local tensors give rise to the same state for all system sizes in 2D. Therefore, if there is a local characterization of such fact must be an uncomputable (and hence useless) one.  

The big question then is to fill the gap in between these two extremes points: the true but rather incomplete normal case and the undecidable general case:

\begin{question}
Give a Fundamental Theorem in 2D (and higher dimensions) for the largest possible family of PEPS.
\end{question}

The relation between the tensors $\mathcal{A}$ and $\mathcal{B}$ has been investigated so far from the equality of their defining PEPS, nevertheless other conditions can be considered.  One of those could be the approximability of two PEPS in the thermodynamic limit:
\begin{question}
Given $\mathcal{A}$ and $\mathcal{B}$ such that there exists an $\epsilon >0$ and a system size $L_0$ such that for all $L \ge L_0$ 
$$\| \ket{\Psi_{\mathcal{A}}} -\ket{\Psi_{\mathcal{B}}}\|_2\le \epsilon,$$
is there a local relation between both tensors?
\end{question}
In contrast to previous questions, here there are no known results; one first step would be answering the question for normal PEPS.

\section{Miscellanea}

There are many other relevant questions about PEPS that were not formulated in the previous sections due to the need of introducing too specialized prerequisites. In this section we will list a selection of those, with the hope that researchers in the corresponding fields could be attracted to such problems:

\paragraph{Machine Learning.} MPS (and other Tensor Networks such as MERA) have been successfully used numerically in the context of Supervised Machine Learning (ML)  \cite{Stoudenmire}. They lack however an in-depth theoretical analysis. A concrete (relevant) question is the following: 

\begin{question}
Can one write the Rademacher complexity or the Vapnik-Chervonenkis (VC)-dimension for such ML algorithms as a function of the bond dimension?
\end{question}
 
 \paragraph{Computational Complexity.} Part of the difficulty of dealing with PEPS is that, as we saw before, they can encode hard (even undecidable) problems. For some type of problems concerning PEPS, the exact complexity class is known \cite{Schuch07}.
 In \cite{uncomputable-PEPS} it is shown that zero-testing in 2D PEPS is a central question to understand their fundamental limitations and the NP-hardness of that problem is proven (see also \cite{Gharibian}).
 
\begin{question}
Which is the exact complexity class for 2D PEPS zero-testing?
\end{question} 
 
 \paragraph{Topological complexity} The complexity of a state (in particular a PEPS) can also be measured in an operational way by the depth of the quantum circuit required to construct it from a different (usually simpler) state. Indeed, fast (meaning low-depth) convertibility in both directions between different states is the quantum-information-like definition for two states to belong to the same quantum phase (see \cite{Coser} for an in-depth discussion on that). One would expect however that one can always reduce complexity fast. Making this statement rigorous for topologically ordered phases boils down to find low-depth circuits of (noisy) gates that implement dynamically the notion of anyon condensation.  The formal question becomes (see \cite{Coser} for the necessary notions and definitions):
 
 \begin{question}
Is there a low-depth noisy quantum circuit that maps the quantum double phase associated to a finite group $G$ to the one associated to a normal subgroup $H$?. 
\end{question} 

\paragraph{Quantum Cellular Automata.} Quantum Cellular Automata (QCA) are unitary evolutions on a lattice that have a finite propagation cone \cite{QCA-Werner}. By means of the Lieb-Robinson bounds they can be seen as discrete analogues of time-evolutions of locally interacting systems. In \cite{QCA} (see also \cite{QCA-chen}) it is shown that 1D translationally invariant QCAs correspond exactly with the set of Matrix Product Unitaries MPU (MPOS that are unitary for all system size). This opens the possibility to combine techniques from MPS and QCAs in order to classify the different QCAs up to continuous deformations, as illustrated in \cite{QCA} and \cite{QCA-onsite}.  The question is:

\begin{question}
Which is the exact relation between PEPS and QCA in 2D and higher dimensions? 
\end{question} 

See \cite{Haah} for recent work in this direction.
\section*{Acknowledgments}

J.G.R. and D.P.G. acknowledge financial support from MINECO (grant MTM2014- 54240-P), from Comunidad de Madrid (grant QUITEMAD+- CM, ref. S2013/ICE-2801), and the European Research Council (ERC) under the European Union's Horizon 2020 research and innovation programme (grant agreement No 648913). J.I.C acknowledges funding through ERC Grant QUENOCOBA, ERC-2016-ADG (Grant No. 742102). This work has been partially supported by ICMAT Severo Ochoa project SEV-2015-0554 (MINECO).

\end{document}